\documentstyle[aps,prl]{revtex}
\begin{document}
\draft
\twocolumn[
\title{Collective Behavior of Coupled Chaotic Maps}
\author{M. G. Cosenza}
\address{Postgrado de Astrofisica, Facultad de Ciencias,
Universidad de Los Andes, \\A. Postal 26, La Hechicera, M\'erida 5251,
Venezuela}
\date{to appear Phys. Lett. A, 1995}
\maketitle
\begin{abstract}
\widetext
\vspace{-0.5cm}
The collective behavior of a coupled map lattice having {\it unbounded} chaotic
local dynamics is investigated through the properties of its mean field.
The presence of unstable periodic orbits in the local maps determines
the emergence of nontrivial collective behavior.
Windows of collective period-two states are found in parameter space.
\narrowtext
\end{abstract}
]

Coupled map lattices (CML) have provided fruitful models for the study of
many spatiotemporal processes in a variety of contexts \cite{r1}.
Recently, there has been great interest in the use of CML models in
the investigation of global
behaviors of networks of chaotic elements \cite{Chate,Bohr,Kaneko}.
Such collective
phenomena
have many implications ranging from the fundamentals of statistical
mechanics \cite{Grin} to
biological information processing, and even possible practical applications
\cite{Kan2}.

In this letter, we study the collective behavior of
the one-dimensional CML system
\begin{eqnarray}
\label{cml}
x_{t+1}(i)=&f(x_t(i))+\gamma[f(x_t(i-1))+f(x_t(i+1)) \nonumber \\
           & -2f(x_t(i))] ,
\end{eqnarray}
where $t$ is a discrete time, $i$ labels the lattice sites $(i=1,\ldots,,N)$,
$\gamma$ is a parameter measuring the diffusive coupling strength between
neighboring sites, and $f$ is a map describing the local dynamics.
Periodic boundary conditions are assumed.

The collective behavior of the lattice can be monitored through
the instantaneous mean field or {\it activity} of the network, defined
as the space average of the local variables $x(i)$ at time $t$
\cite{Chate}:
\begin{equation}
S_t=\frac{1}{N} \sum_{i=1}^{N}x_t(i).
\label{mean}
\end{equation}
A wide variety of local map functions $f$ have been employed in CML
models depending on the particular application. Usually, bounded maps
belonging to some universality class are considered to be the source
of local chaos in the study of collective spatiotemporal dynamics
in CMLs with local \cite{Chate,Bohr} or global couplings
\cite{Kaneko}. Here we investigate the collective behavior
of the lattice described by (\ref{cml}) having unbounded local dynamics
such as the logarithmic map
\cite{Kawabe}
\begin{equation}
\label{local}
f(x)=b+\ln|x|.
\end{equation}
This map posseses no maximum or minimum and its Schwarzian derivative is
always positive.
Two stable fixed points satisfying $f(x^*)=x^*$ exist for this map:
$x_1^* < -1$, for $b<-1$, which becomes unstable at $b=-1$;
and $x_2^*>1$, for $b>1$, which arises from a tangent bifurcation
at $b=1$.
Chaos occurs in the parameter region $b \in [-1,1]$.
There are no gaps separating chaotic bands at any given value of
$b \in [-1,1]$ and no periodic windows appear in any subinterval of $b$ in
this range \cite{Kawabe}.
These special features, contrasting with the universal properties of
the commonly used local maps,
make maps of the type (\ref{local}) an alternative tool for exploring
the essential requirements for
the emergence of collective behavior in CMLs and the generality of this
phenomenon.

The dynamics of the CML~(\ref{cml}) can be written in vector form as
\begin{equation}
\label{matrix}
{\bf x}_{t+1}={\bf f}({\bf x}_t)+\gamma {\bf M}{\bf f}({\bf x}_t) ,
\end{equation}
where the N-dimensional vectors ${\bf x}_t$ and ${\bf f}({\bf x}_t)$
have components $[{\bf x}_t]_i=x_t(i)$ and
$[{\bf f}({\bf x}_t)]_i=f(x_t(i))$, respectively; and ${\bf M}$ is a
$N \times N$ tridiagonal, symmetric matrix expressing diffusive
coupling among the components $[{\bf x}_t]_i$. The non-zero components
of ${\bf M}$
are $M_{ii}=-2$; $M_{ij}=1$ $(i=j\pm1)$.

With the local map (\ref{local}), the stable, spatially homogeneous,
stationary states $x_t(i)=x_1^*$, for $b<-1$;
or $x_t(i)=x_2^*$, for $b>1$, $\forall i$, are
possible for the system~(\ref{cml}). The linear stability analysis of
these states leads to the bifurcation conditions
\begin{equation}
\label{bif}
(1+\gamma\mu_k)f'(x_{1,2}^{*})=\pm 1,
\end{equation}
where $\{\mu_k: \; k=1,\ldots,N\}$ is the set of eigenvalues of the
coupling matrix ${\bf M}$ which satisfy $\mu_k \in [-4,0]$ \cite{Waller}.
Equations~(\ref{bif}) yield boundary curves in the parameter plane
$(\gamma,b)$ which determine where each homogeneous, stationary state
can be observed.
Figure~1 shows the first stability boundaries for each state, corresponding
to $\mu_k=-4$.
The state $x_t(i)=x_1^*$, $\forall i$, is stable for parameter values in
the region enclosed by the r.h.s.
$\pm1$ boundaries corresponding to $x_1^*$ in Eq.~(\ref{bif}), with $b<-1$.
Similarly, the state $x_t(i)=x_2^*$, $\forall i$,
is stable inside the parameter region
bounded by the r.h.s. $\pm1$ curves associated to $x_2^*$, with $b>1$.
Within these regions of stable homogeneous stationary states, the
asymptotic values of the mean field are $S_t=x_1^*$ and $S_t=x_2^*$,
respectively. The crossing of either boundary signals the appearance
of a spatially inhomogeneous state which should be reflected in a
dispersion of $S_t$.

When the value of the parameter $b$ is in the range $[-1,1]$,
corresponding to local
chaotic dynamics, the asymptotic collective behavior of the lattice, as given
by $S_{t \rightarrow \infty}$, reveals the existence of global
low-dimensional periodic attractors, subjected to fluctuations of intrinsic
statistical origin.
Figure~2 shows the bifurcation diagram of the asymptotic mean field $S_t$
as a function of the local parameter $b$. The quantity $S_t$ was calculated
at each time step during a run starting from random initial conditions
uniformly distributed on the interval [-8,4] at each site and for each value
of $b$, after discarding the transients. The coupling strength was fixed
at $\gamma=0.5$, corresponding to a ``totalistic" coupling of the ``game
of life" type \cite{Chate}. Similar bifurcation diagrams can be obtained
for other fixed values of the coupling parameter.

For $b<-1$ or $b>1$, the asymptotic mean field is identical to the
values of the corresponding fixed points of the single map
(\ref{local}) in these ranges of $b$,
as expected from the stability diagram of the
homogeneous stationary states in Fig.~1. In the region $b \in [-1,1]$,
figure~2 shows a pitchfork bifurcation at a value $b_c \simeq -0.52$
from a collective
fixed point state (a state for which the time series of $S_t$
manifestly displays statistical fluctuations around a single value) to
a collective period-two state (a state for which the time series of
$S_t$ alternatively varies between the corresponding neighborhoods of
two separated values).
The small vertical segments seen in Fig.~2
can be interpreted as the amplitude of the fluctuations of the mean
field about the corresponding global attractor at given parameter values. The
fluctuations around the global stationary or the period-two attractors
are due to the fact that the local variables behave chaotically, as can
be attested by the time series of any site and by the existence of a
positive largest Lyapunov exponent.

The appearance of a period-two collective state in the one-dimensional
lattice (\ref{cml}) is related to the fact that the iterates of the
local map (\ref{local}) move alternatively from values above the unstable
fixed point $x_1^*$ to values bellow this point in the interval
$b \in [-1,b_c]$, even though there are no separated chaotic bands
\cite{Kawabe}.
The iterates behave chaotically within each side of the unstable fixed
point $x_1^*$,
but {\it periodically} about it.
The unstable fixed point $x_1^{*}$ establishes
a ``symmetry" line around which the iterates oscillate with period two.
As a comparison,  such effect does not occur in the unbounded
map $g(x)=a-1/x$, which lacks unstable periodic orbits for
$a \in [-2,2]$; consequently, no collective periodic behavior
emerges in a coupled lattice of these maps at that
range of the parameter $a$ \cite{MC}.

Coupling induces synchronization of the array
in the range $b \in [-1,b_c]$, in the sense that iterates of
the local chaotic sites
tend to move together above and bellow
the unstable fixed point $x_1^{*}$ at alternate time steps.
The individual local values may be
different within each of these two regions at a given instant. Figure~3(a)
presents the asymptotic mean field $S_t$ as a function of the coupling strength
$\gamma$, keeping constant $b=-0.8$. When $\gamma$ is close to zero, and
starting each time from random initial conditions, $S_t$
fluctuates about an average value of the desynchronized local maps.
There exists a critical value of the coupling at
$\gamma_c \simeq 0.01$ where the transition from a statistical average
to a period-two collective state takes place. The periodic behavior of $S_t$
remains practically unchanged until the coupling reaches a second critical
value $\gamma'_c \simeq 0.53$ at which synchronization is again lost.
Figure~3(b) shows the asymptotic behavior of an arbitrary individual site
simultaneously monitored as a function of the coupling.
The value of the unstable fixed point $x_1^*$ corresponding to $b=-0.8$
is indicated by a horizontal line.
Coupling enhances
the separation of alternate iterates about $x_1^*$, creating
a gap between them for $\gamma \in [\gamma_c,\gamma'_c]$.
The mean field reflects this separation as well as
the synchronization of the lattice
produced by the coupling.

Figure~4 shows the dependence of the asymptotic $S_t$ on the size of
the lattice $N$, with the other parameters held fixed $(\gamma=0.5,
b=-0.8)$. There is a rather small critical size $N_c \simeq 10$ at which the
global period-two attractor distinctively emerges. As in Fig~2,
the vertical segments
represent fluctuations around the global period-two attractor and they
correspond to even or odd steps of the asymptotic time series of
$S_t$, respectively. The variance of either subset of steps of the time series
of $S_t$ decreases as $N^{-1}$, as expected. However, the variance of the
mean field itself, for the chosen parameters,
tends to a constant value for large enough  $N$.
In the limit $N \rightarrow \infty$,
the global period-two orbit will have the form
$S_t: \, \ldots s_1,s_2,s_1,s_2 \ldots$. The time average of the
mean field will be $\bar{S}=(s_1+s_2)/2$, while the variance $\sigma$ will
yield
\begin{equation}
\sigma=\frac{1}{T}\sum_{t=1}^{T}(S_t-\bar{S})^2=
\frac{1}{T}\sum_{t=1}^{T}\frac{(s_1-s_2)^{2}}{4} .
\end{equation}
For the parameters of Fig.~4, $s_1\simeq -2.55, s_2 \simeq 0.08$.
The limiting value $\sigma \simeq 1.72$ is approached
for $ N \simeq 10^3$.
The existence of a saturation value for the variance of the
mean field
characterizes the emergence of nontrivial ({\em i.e.}
periodic, quasiperiodic, or chaotic)
collective behavior. This phenomenon has been called ``violation of the
law of large numbers" in the context of globally coupled maps
belonging to some universality class (tent, quadratic, or circle maps)
\cite{Kaneko,Kurths}.

The onset of a periodic collective state at some values
of the parameters of the system is reminiscent of a phase transition,
because it corresponds to abrupt changes in the statistical
properties of the lattice, as described by $S_{t \rightarrow \infty}$,
a quantity which acts as an order parameter.
We have found windows of global periodic behavior in the space of
parameters.
Periodic collective states have been observed in higher dimensional
lattices of locally chaotic coupled maps. In
those cases, the collective behavior consists in statistical cycling
of the mean field (or the probability density)
among chaotic bands of the local
one-hump maps \cite{Chate,Bohr,Mackey}.
This paper shows that high space dimension, large system size,
strong coupling, bounded
iterates, gap-separated chaotic bands, or the existence of
periodic windows in the local dynamics, are not essential
requirements for the emergence of nontrivial collective behavior.
We have found that coupling can synchronize the cycling of the
chaotic iterates and enhances their separation around unstable
periodic orbits of the local maps in some parameter ranges
giving rise to periodic global behavior in a lattice.
Our results suggest that the emergence of nontrivial collective
behavior should be a rather
generic phenomenon in deterministic systems of coupled chaotic
elements, where unstable periodic orbits are always present.

The author thanks Prof. A. Parravano for useful discussions. This work has
been supported by grant C-658-94-05-B from Consejo de Desarrollo
Cient\'{\i}fico,
Human\'{\i}stico y Tecnol\'ogico of Universidad de Los Andes, Venezuela.

\begin{figure}
\caption{
Stability boundaries for the homogeneous stationary
states of the
CML~(\protect\ref{cml}).
The labels on each curve identify the first
stability boundaries with $\pm1$
in Eq.~(\protect\ref{bif}) for each state, corresponding to $\mu_k=-4$.
The vertical lines at $b=-1$ and $b=1$ correspond to $\mu_k=0$. Curves
associated with other eigenvalues lie outside the regions enclosed
by the boundaries shown. The states $S_t=x_1^{*}$ and $S_t=x_2^{*}$ are
stable inside the indicated regions.}
\end{figure}
\begin{figure}
\caption{Bifurcation diagram of the asymptotic $S_t$, as a function
of the local parameter $b$. Coupling is fixed at $\gamma=0.5$.
For each value of $b$, 100 iterates are shown, after discarding $5000$
transients.  Lattice
size is $N=10^5$. For $b<-1$ and $b>1$, $S_t$ corresponds to the
values of the homogeneous
stationary states $x_1^{*}$ and $x_2^{*}$, respectively.}
\end{figure}
\begin{figure}
\caption{a) $S_t$ as a function of the coupling parameter
$\gamma$. Local parameter is fixed at $b=-0.8$. For each value of $\gamma$,
100 iterates are shown, after discarding 5000 transients. Lattice size is
$N=10^5$. b) Asymptotic iterates of site $i=100$, for the same parameters as
in a). The horizontal line corresponds to the value of the unstable $x_1^*$
for the given $b$.}
\end{figure}
\begin{figure}
\caption{Asymptotic mean field $S_t$ as a function of lattice size $N$.
Fixed parameters
are $\gamma=0.5$; $b=-0.8$.}
\end{figure}
\end{document}